# A liquid film can survive without being bulk-like: Rupture and mechanical isotropy set different thickness limits in Lennard-Jones films

José L. Altamirano-Corona [1], Alfonso Lemus-Solorio [1], Francisco Villanueva-Mejia [2], Mariana Ramos-Estrada [1] and José L. Rivera [3,*]

[1] *Facultad de Ingeniería Química, Universidad Michoacana de San Nicolás de Hidalgo, Morelia 58000, Michoacán, México*

[2] *Instituto Tecnológico de Pabellón de Arteaga, Carretera a la estación de Rincón de Romos, km 1, Pabellón de Arteaga 20670, Aguascalientes, México*

[3] *Facultad de Ciencias Físico-Matemáticas, Universidad Michoacana de San Nicolás de Hidalgo, Morelia 58000, Michoacán, México*

How thick must a liquid film be before its middle behaves like ordinary bulk liquid? Free-standing films are the standard geometry for computing surface properties by molecular simulation and the simplest model of soap and foam films, coatings and polymer layers, yet the question is seldom answered quantitatively. We answer it for Lennard-Jones films by comparing two criteria: whether the film survives without rupturing, and whether the lateral pressure at its center has relaxed to the bulk value. Covering three interaction ranges (cutoffs $r_c = 3.5$–$7.5\sigma$), three temperatures and lateral system sizes from 20 to 100σ, we find that the two criteria give different thicknesses and respond to different parameters. Films thinner than 6–7σ (about 2 nm for argon) rupture within $10^5$ τ (≈0.2 μs); this kinetic limit hardly depends on the interaction range but grows with the lateral size of the system, and close to it rupture is an activated event. The center becomes mechanically bulk-like only at a larger thickness $h^*$, where the tangential anisotropy computed with the Irving–Kirkwood definition falls below 0.003 $\varepsilon/\sigma^3$; $h^*$ grows from ≈11σ (3.8 nm) to ≈15σ (5 nm) as the cutoff increases from 3.5 to 7.5σ, but it is nearly independent of temperature and of lateral size. Between the two limits films are long-lived yet perturbed throughout. Neither the surface tension, which shows no thickness dependence, nor the central density, whose relative deficit is about 16 times smaller than the relative anisotropy, reveals this perturbation: the stress detects confinement before the structure does.

*Contact author: jlrivera@umich.mx

## I. INTRODUCTION

Planar liquid films that are free on both sides are the workhorse geometry for simulating vapor–liquid interfaces. A slab of liquid placed in an elongated periodic box develops two interfaces, and the surface tension is computed from the difference between the pressures normal and parallel to them [1–8]. Such films are also the simplest model of thin films that matter well beyond simulation: soap and foam films, the liquid lamellae between bubbles, molten polymer layers, and lubricant and coating films. All these uses rest on an assumption that is rarely checked: that the film is thick enough for its two surfaces to be independent, with ordinary bulk liquid in between. If the film is too thin, its middle is stretched sideways, the two surfaces feel each other, and properties measured on it depend on thickness in a way that is easy to mistake for a property of a single surface [9–11].

Two questions follow. The first is mechanical: at what thickness does the pressure tensor at the film center become isotropic within a stated tolerance? We proposed this isotropy criterion for Lennard-Jones (LJ) films [12] and later applied it to molten polyethylene layers, where the mechanically defined critical thickness exceeded the thickness at which films visibly rupture by a factor of two to three [13]. The second question is kinetic: for how long does a thin film survive before it perforates and transforms into a cylinder or a drop? Rupture of thin free films has a long history in colloid science, where it is usually discussed in terms of the growth of thickness fluctuations driven by the disjoining pressure [14–16], and molecular simulations have shown that nanoscale films rupture and dewet on accessible time scales [17,18]. Deterministic thin-film models describe the dewetting of such films well [19], but at nanometer thicknesses thermal fluctuations of the surfaces accelerate the instability [20–22].

The mechanical question has a technical complication. The pressure at a point inside an interface is not unique: it depends on how the force between two molecules is assigned to the points of space between them (the integration contour), a choice that makes no difference in a bulk liquid but does near an interface [23–27]. The tangential component computed from per-atom virials (PAV) [28], which corresponds to the Harasima contour [29], differs from the Irving–Kirkwood (IK) one [30] by the derivative of a contour-dependent term, so that thresholds based on the PAV profile are biased in thin films. In a companion paper we derived and validated an estimator that recovers the IK tangential pressure from PAV data [31]; here we use the IK tensor directly and use the estimator only as an internal consistency check.

In this work we compare the two thresholds systematically. We measure the central anisotropy and the rupture behavior of LJ films as functions of thickness, interaction cutoff $r_c$, temperature and lateral box size $b_t$. We also compare the central anisotropy with the two quantities that are usually checked to decide whether a film is thick enough, the density at the film center and the surface tension. The main result is that the two criteria respond differently to these parameters: the mechanical threshold $h^*$ depends on the range of the potential but not on $b_t$, whereas the kinetic threshold is insensitive to the cutoff but moves with $b_t$. Between the two thresholds lies a range of thicknesses in which films are long-lived but mechanically perturbed, which is precisely the range in which they are most often used. Finally, we relate the central anisotropy to the disjoining pressure of film thermodynamics and use that relation to explain why the surface tension does not reveal it. Figure 1 summarizes these results; for argon, the perturbed range spans roughly 2.4 to 5 nm.

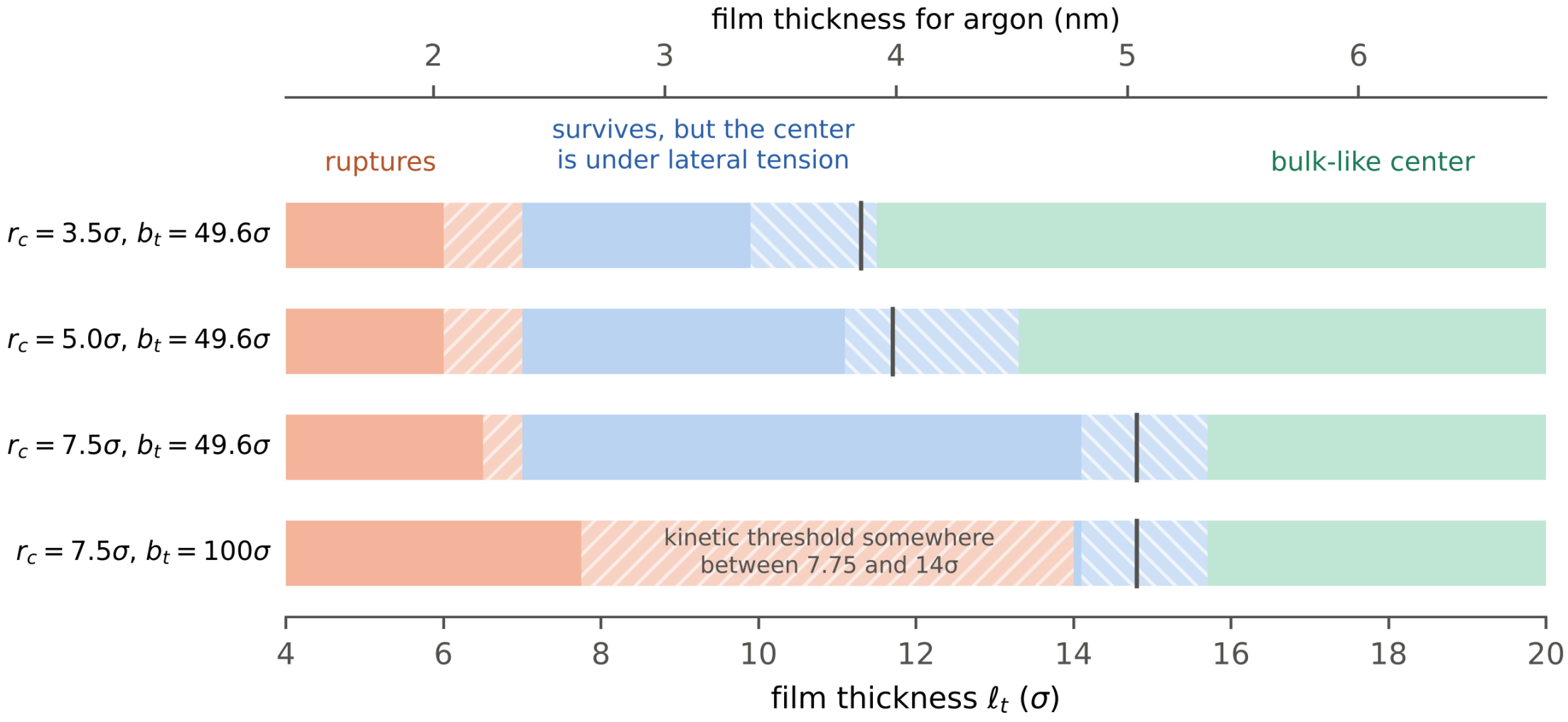

FIG. 1. Summary of the two stability limits of LJ films at $T^* = 0.72$. Below the kinetic limit films rupture within $10^5$ $\tau$ (≈0.2 μs for argon); between the kinetic limit and the mechanical threshold $h^*$ (vertical lines, defined by a central tangential anisotropy of 0.003 $\varepsilon/\sigma^3$) they survive but the liquid at their center is under lateral tension; above $h^*$ the center is bulk-like. Hatched regions show the uncertainty of each limit; at $b_t = 100\sigma$ the kinetic limit lies somewhere between 7.75σ, which ruptured, and 14σ, which survived. The top axis converts thickness to nanometers with σ = 0.34 nm.

## II. METHODS

### A. Model and simulations

Atoms interact through the LJ potential [32] with energy ε and size σ, truncated without shifting at $r_c$ = 3.5, 5.0 or 7.5σ. Because the forces are cut at $r_c$ and no impulsive term from the discontinuity of the potential is added, the dynamics and the pressure tensor are those of the truncated-and-shifted (STS) model in the terminology of Trokhymchuk and Alejandre [7]; the shift itself would change only the potential energy. The way the potential is truncated affects the coexistence properties and the surface tension of LJ fluids [7,33,34], which is why all results are reported together with $r_c$. Reduced units are used throughout; time is in units of $\tau = \sigma(m/\varepsilon)^{1/2}$. For orientation, with the argon parameters σ = 0.34 nm and $\varepsilon/k_B$ = 120 K [35], τ ≈ 2.2 ps, the temperatures $T^*$ = 0.72–0.90 correspond to 86–108 K, a pressure of 0.003 $\varepsilon/\sigma^3$ is ≈0.13 MPa, and a surface tension of 1 $\varepsilon/\sigma^2$ is ≈14 mN/m. Simulations were run with LAMMPS [36] in the canonical ensemble with a Nosé–Hoover thermostat [37,38] and a time step of 0.005τ. Each film lies in the $xy$ plane of a periodic cell of lateral size $b_t \times b_t$ and height $L_z$ = 48σ, with the film normal along $z$;

the center of mass of the film is kept at a fixed position. Films were built from a lattice slab, melted, and equilibrated for $2 \times 10^7$ steps ($10^5$ τ). Production runs of $2 \times 10^6$ steps ($10^4$ τ) recorded the IK and PAV pressure-tensor profiles in bins of 0.5σ along $z$, together with the density profile.

Five groups of systems were simulated (Table I). The thickness series at $T^* = 0.72$ and $r_c = 7.5\sigma$ uses $b_t = 49.6\sigma$ and thicknesses from 7 to 25σ. The cutoff series repeats it at $r_c$ = 3.5 and 5.0σ, and the temperature series at $T^* = 0.80$ and 0.90. The lateral series keeps the number of atoms per unit area of selected films fixed and varies $b_t$ from 20 to 100σ. Independent replicas, started with different initial velocities, were run for the films nearest to each threshold. A separate set of runs without pressure analysis was used to locate the rupture threshold.

TABLE I. Simulated systems. $\ell_t$ is the distance between the two half-density points at the interfaces. Systems listed under "Rupture only" were run without pressure analysis; films of the other series that ruptured are listed there as well. Full lists are given in the Supplemental Material [39], Tables S1–S6.

| Series | $T^*$ | $r_c$ (σ) | $b_t$ (σ) | $\ell_t$ (σ) | Systems | Trajectories |
|---|---|---|---|---|---|---|
| Thickness | 0.72 | 7.5 | 49.6 | 7.0–25.0 | 7 | 15 |
| Cutoff | 0.72 | 3.5, 5.0 | 49.6 | 7.0–17.6 | 13 | 17 |
| Temperature | 0.80, 0.90 | 7.5 | 49.6 | 8.1–17.6 | 10 | 12 |
| Lateral size | 0.72 | 3.5, 7.5 | 20–100 | 7.0–14.1 | 10 | 16 |
| Rupture only | 0.72 | 3.5, 5.0, 7.5 | 20–100 | 5.5–7.75 | 12 | 32 |
| Total | | | | | 52 | 92 |

### B. Pressure tensor, thickness and surface tension

The film thickness $\ell_t$ is the distance between the two points where the density equals the mean of the liquid and vapor densities, located by linear interpolation; the film center $z_c$ is their midpoint. The surface tension is

$$\gamma = \tfrac{1}{2} \int [P_N(z) - P_T(z)]\, dz \qquad (1)$$

which is the same for the IK and PAV contours because the two tangential profiles differ by the derivative of a function that vanishes in the bulk phases [31]. The quantity of interest here is the tangential anisotropy at the film center,

$$\Delta P_T = P_{vap} - \langle P_T^{IK} \rangle, \quad |z - z_c| \leq \sigma \qquad (2)$$

where $P_{vap}$ is the normal IK pressure averaged over the vapor bins and the average of the tangential IK pressure runs over a window of fixed width 2σ centered on $z_c$, with fractional weights for the bins cut by the window edges. In a thick film the center is bulk liquid and $\Delta P_T = 0$; a positive value means that the liquid at the center is under lateral tension. As a consistency check, the IK profile estimated from PAV data with the closure of Ref. [31] reproduces the measured one in every system (Supplemental Material [39], Sec. S9).

Statistical errors are standard errors over 200 consecutive blocks of the production run; where replicas exist we report their mean and the standard error between replicas, which is larger than the block error in the smallest boxes (Sec. S9).

### C. Mechanical threshold

We define the mechanical threshold $h^*$ as the thickness at which $\Delta P_T$ falls below a tolerance of 0.003 $\varepsilon/\sigma^3$, which is about twice the statistical error of a single film at $b_t = 49.6\sigma$. $h^*$ is obtained by interpolating ln $\Delta P_T$ linearly between the two thicknesses that bracket the tolerance (linearly in $\Delta P_T$ when the thicker of the two is negative). Its uncertainty is the 16–84% interval of $2 \times 10^4$ Monte Carlo resamplings of all $\Delta P_T$ values within their errors. Because $\Delta P_T$ decays roughly exponentially with a length $L$, changing the tolerance by a factor of 2 moves $h^*$ by about $L$ ln 2, that is, by 1–2σ; the tolerance is therefore part of the definition and is stated with every value.

### D. Rupture detection

Rupture was detected from the potential energy per atom, whose drop marks the loss of surface area, and from the shape of stored configurations, classified as film, perforated film, cylinder or drop (Fig. 2; details and an example in the Supplemental Material [39], Sec. S7 and Fig. S1). A film is counted as kinetically stable when it remains a film for the full $2 \times 10^7$-step equilibration ($10^5$ τ) in all replicas.

### E. Analysis software and use of AI tools

The profiles were analyzed with Python scripts (gauge_291026b.py for the pressure profiles, thresholds and surface tensions; ruptura_021126.py for the detection of rupture), which are available from the author. These scripts, and those that produce the figures from the simulation data, were written with the assistance of an AI tool (Claude, Anthropic, model Fable, version 5.2), following specifications given by the author. They were validated by tests on synthetic profiles with known answers and, on the simulation data, by the agreement between the estimated and the directly measured IK profiles and by the comparison of block and replica errors (Supplemental Material [39], Sec. S9). The author ran all the simulations and analyses. No figure was generated by a generative model.

## III. RESULTS

### A. Structure of thin and thick films

Figure 2 shows the systems studied. A stable film is a flat slab with rough, fluctuating surfaces and a few atoms in the vapor; when it ruptures, a hole opens and the liquid reorganizes into a cylinder and then a drop. Figure 3 compares the density and tangential-pressure profiles of three films at $T^* = 0.72$ and $r_c = 7.5\sigma$. The density of the 7σ film still reaches a central plateau within 1.2% of the bulk value. The tangential pressure is negative in the interfacial region, with a minimum of −0.44 to −0.46 $\varepsilon/\sigma^3$ on each side, and in the 25σ film it returns to the vapor pressure, $P_{vap}$, over a wide central region. In the thinner films it does not: between the two minima $P_T$ stays below $P_{vap}$, and the difference at the center, $\Delta P_T$, is 0.10 $\varepsilon/\sigma^3$ at 7σ and 0.01 $\varepsilon/\sigma^3$ at 11σ [Fig. 3(c)]. The per-atom-virial profile reproduces the shape but underestimates the central anisotropy (0.077 against 0.105 $\varepsilon/\sigma^3$ at 7σ), which is why the IK contour is needed for the threshold [31]. The normal pressure, as required by mechanical equilibrium, is flat and equal to $P_{vap}$ within 0.003 $\varepsilon/\sigma^3$ in all films.

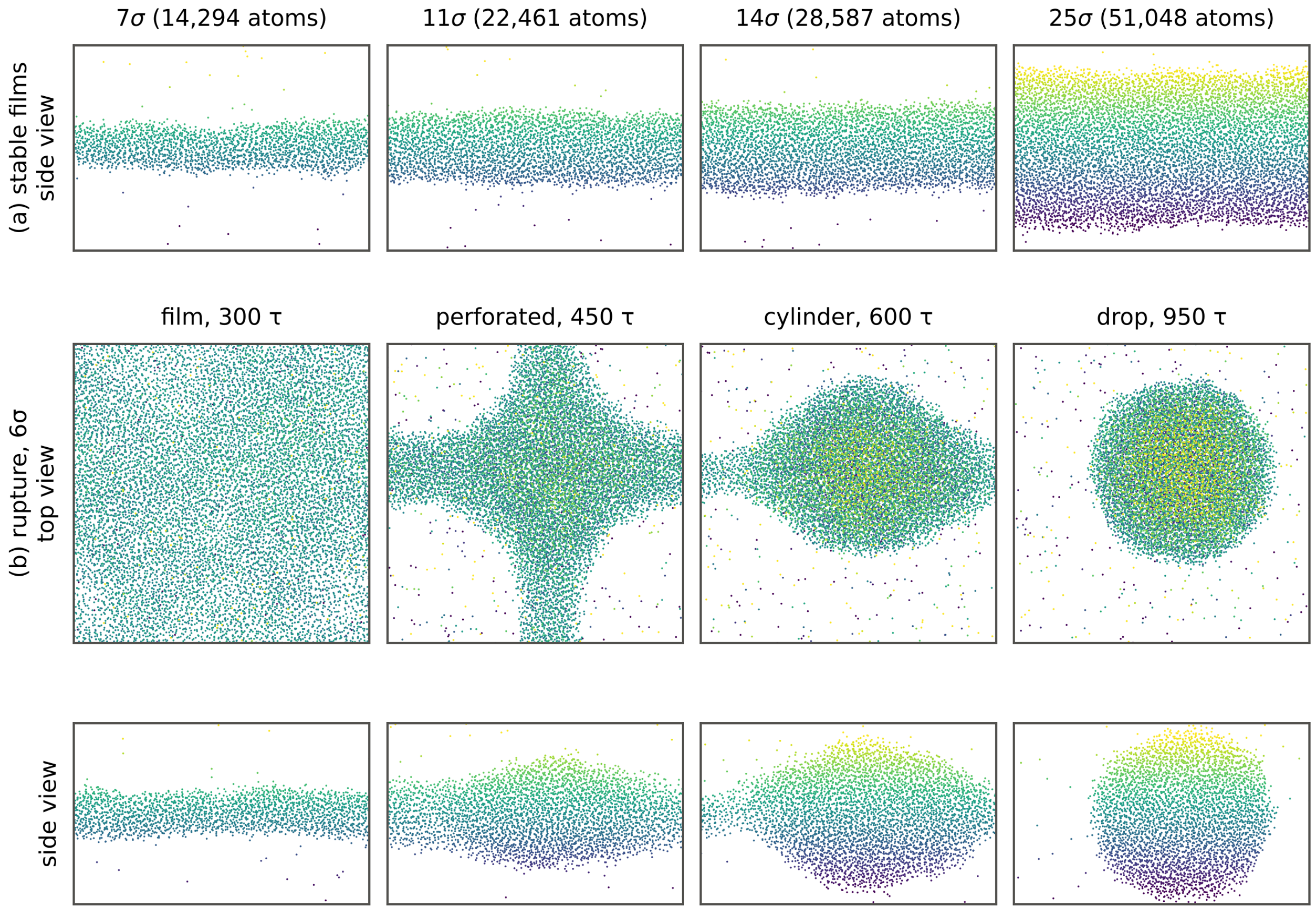

FIG. 2. Snapshots of LJ films at $T^* = 0.72$ and $r_c = 7.5\sigma$ ($b_t = 49.6\sigma$); atoms are colored by height. (a) Side views (slices 8σ thick) of stable films of 7, 11, 14 and 25σ at the end of their production runs. (b) A film of 12,252 atoms ($\ell_t = 6\sigma$) before and during rupture, seen from above (middle row) and from the side (bottom row).

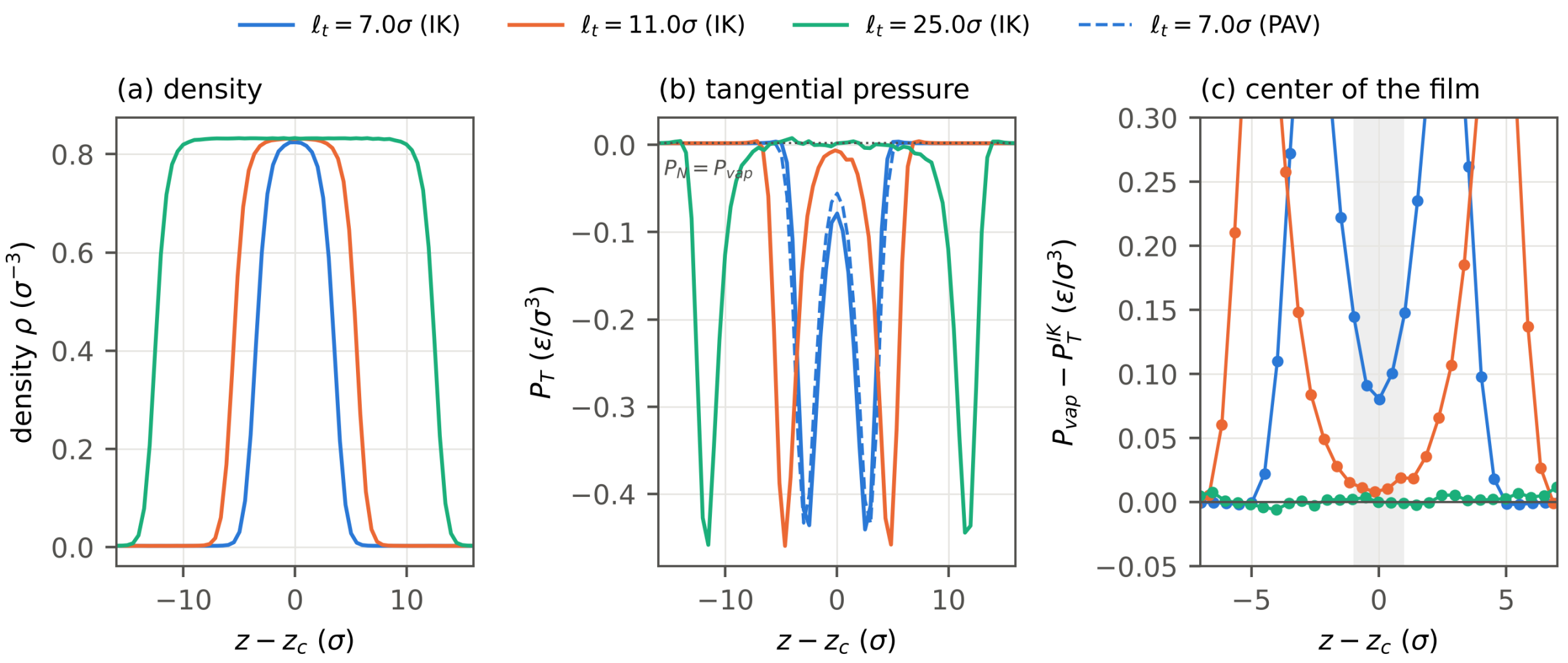

FIG. 3. Profiles of films of 7.0, 11.0 and 25.0σ at $T^* = 0.72$ and $r_c = 7.5\sigma$ ($b_t = 49.6\sigma$), centered on the film center $z_c$. (a) Density. (b) Tangential pressure, IK (solid) and per-atom virial (dashed, 7σ film only); the dotted line is the vapor pressure, equal to the normal pressure. (c) Deficit of the IK tangential pressure with respect to the vapor pressure near the center; the shaded band is the window that defines $\Delta P_T$.

## B. The stress detects confinement before the structure

Figure 4 compares how three observables converge with thickness. The surface tension [Fig. 4(a)] shows no trend: over each series its values scatter within ±0.6% ($r_c = 7.5\sigma$) to ±1.3% (3.5σ) around the mean, with no systematic trend, from 7 to 25σ. The central density [Fig. 4(b)] does approach its asymptotic value from below: at $r_c = 7.5\sigma$ the deficit $1 - \rho_c/\rho_\infty$ is 1.15% at 7σ, 0.71% at 7.75σ and 0.12% at 11σ, and from 14σ on it is below 0.02%, the level of the statistical noise. The central anisotropy [Fig. 4(c)], expressed as a fraction of the kinetic pressure $\rho_c k_B T$, is 17.5%, 11.4% and 1.8% at the same three thicknesses and still 0.64% at 14σ. The two signals are proportional: for every film whose density deficit exceeds 0.1%, at all three cutoffs, the ratio of the relative anisotropy to the relative density deficit lies between 14 and 18, with a mean of about 16. The density deficit is therefore a weak echo of the same perturbation, and it is weak because the liquid is stiff. The isothermal bulk modulus of the bulk LJ liquid with the same potential, obtained from NVT runs at five densities around $\rho_\infty$, is $K_T = 11.8 \pm 0.1$ $\varepsilon/\sigma^3 \approx 20\rho k_B T$ at $T^* = 0.72$ (Supplemental Material [39], Sec. S12); equations of state fitted to the full LJ potential [40,41] were not used because they do not include the truncation. An elastic response of bulk liquid to the central stress of the 7σ film would lower its density by only 0.6% if the mean stress $(2/3)\Delta P_T$ is taken, or 0.9% for the full $\Delta P_T$, against 1.2% measured. The order of magnitude, and hence the weakness of the density signal, follows from the stiffness of the liquid; the excess shows that the center of a 7σ film is not simply a uniformly stretched bulk liquid. Because it is about 16 times smaller in relative terms, the density deficit falls below the noise between 11 and 14σ for $r_c = 7.5\sigma$ and between 8 and 9σ for $r_c = 3.5\sigma$, whereas the anisotropy remains measurable up to the mechanical threshold $h^*$, 2–4σ further. A film whose central density and surface tension look converged can thus still be under measurable lateral tension.

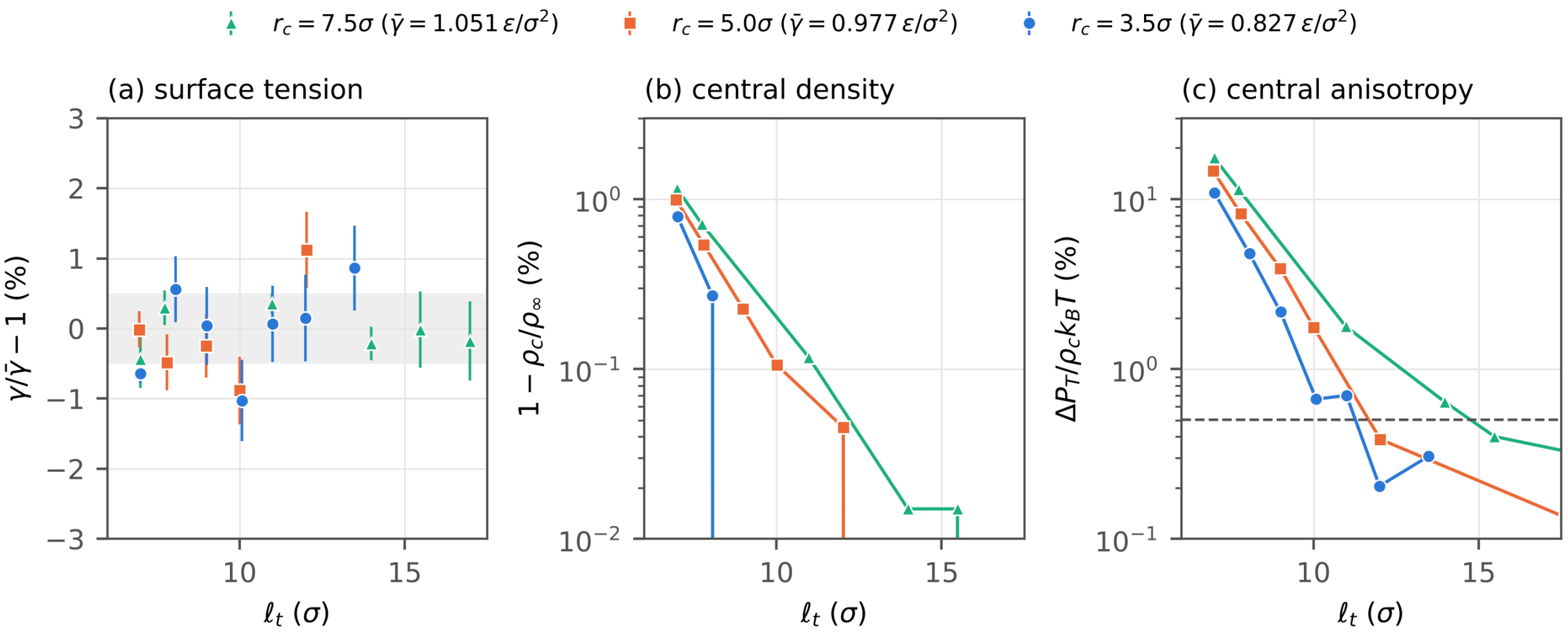

FIG. 4. Convergence with film thickness at $T^* = 0.72$ and $b_t = 49.6\sigma$. (a) Surface tension relative to its mean $\bar{\gamma}$ over each series (values of $\bar{\gamma}$ in the legend, as in Table II); the shaded band is ±0.5%. (b) Deficit of the central density (average over $|z - z_c| \leq \sigma$) relative to the mean of the films thicker than 12σ; points at or below zero are not shown. (c) Central anisotropy relative to the kinetic pressure $\rho_c k_B T$; the dashed line corresponds to the tolerance $0.003\ \varepsilon/\sigma^3$. Replicas are averaged.

## C. Central anisotropy and the mechanical threshold

Figure 5(a) shows $\Delta P_T$ against thickness at $T^* = 0.72$. At $r_c = 7.5\sigma$ the central anisotropy drops from $0.1037 \pm 0.0005\ \varepsilon/\sigma^3$ at $\ell_t = 7.0\sigma$ to $0.0677 \pm 0.0008$ at 7.75σ, $0.0107 \pm 0.0007$ at 11.0σ and $0.0038 \pm 0.0007$ at 14.0σ (means of three replicas), reaching zero within error at 17 and 25σ. The decay is not a single exponential: a fit to all points gives $\chi^2 = 8.5$ for 5 degrees of freedom, whereas separate fits give a decay length $L = 1.75 \pm 0.04\sigma$ up to 11σ and $2.8 \pm 0.5\sigma$ from 11σ on. The slow tail governs the threshold, which is $h^* = 14.8\sigma$ (16–84% interval 14.1–15.7σ). The surface tension does not depend on thickness over the whole series ($\gamma = 1.051 \pm 0.001$, mean and standard error over seven thicknesses), so it gives no warning that the thinner films are mechanically perturbed.

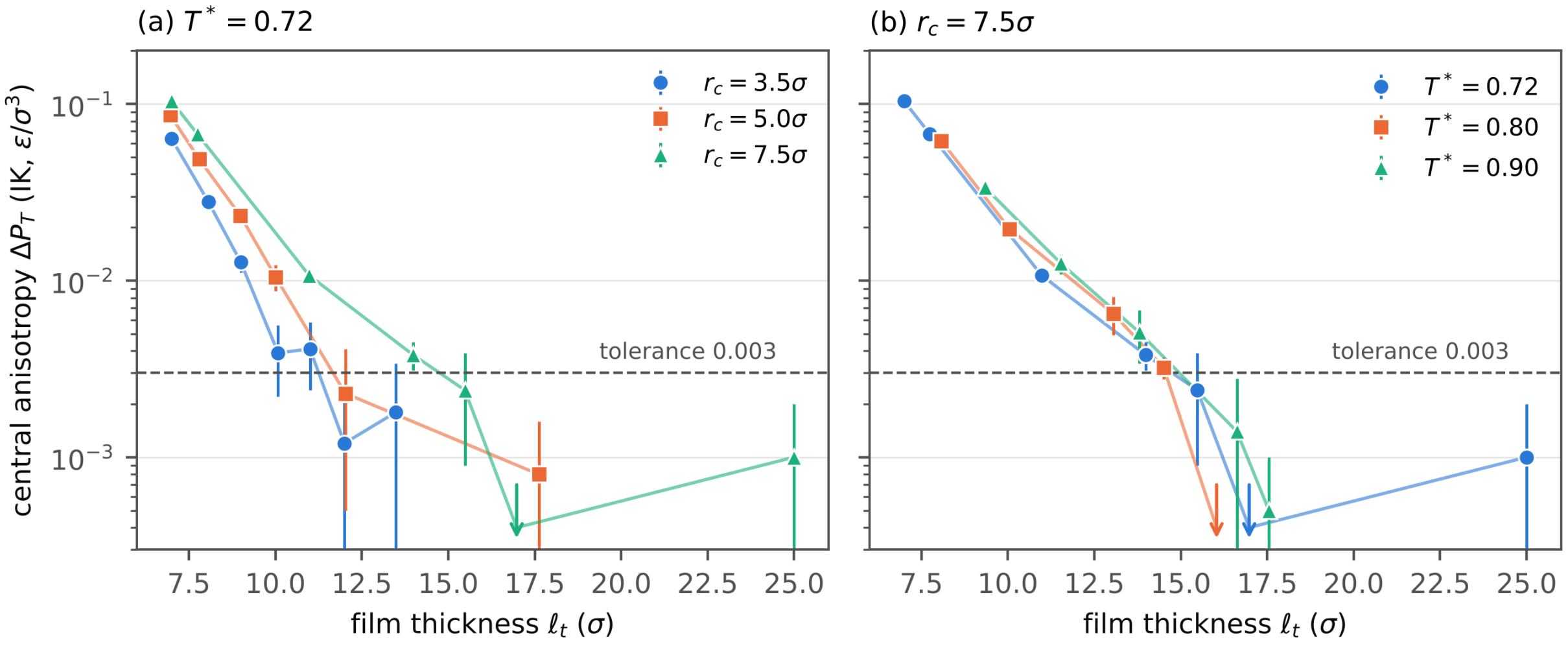

FIG. 5. Central tangential anisotropy $\Delta P_T$ (IK contour) against film thickness, $b_t = 49.6\sigma$. (a) Three cutoffs at $T^* = 0.72$. (b) Three temperatures at $r_c = 7.5\sigma$. Points with replicas are means with the standard error between replicas; arrows mark negative values. The dashed line is the tolerance that defines $h^*$.

### D. Effect of the cutoff

Shortening the cutoff lowers the anisotropy at every thickness and shortens its decay [Fig. 5(a), Table II]. At $\ell_t \approx 7\sigma$, $\Delta P_T$ is $0.063 \pm 0.004$, $0.0868 \pm 0.0013$ and $0.1037 \pm 0.0005$ $\varepsilon/\sigma^3$ for $r_c = 3.5$, 5.0 and 7.5σ, and the decay length in the thin-film regime is 1.26, 1.49 and 1.75σ. The threshold moves accordingly, from $h^* = 11.3\sigma$ at $r_c = 3.5\sigma$ to 11.7σ at 5.0σ and 14.8σ at 7.5σ. $h^*$ therefore grows with the range of the potential, but more slowly than $2r_c$: the ratio $h^*/r_c$ falls from 3.2 at $r_c = 3.5\sigma$ to 2.0 at 7.5σ, so that $h^*$ exceeds $2r_c$ by 60% at the shortest cutoff and nearly equals it at the longest. The anisotropy at the center of a film is thus not simply the direct interaction between molecules near the two surfaces, which would vanish beyond $2r_c$; it also carries the structural perturbation of the interfacial region, whose extent depends only weakly on the cutoff.

TABLE II. Mechanical threshold $h^*$ (tolerance 0.003 $\varepsilon/\sigma^3$, $b_t = 49.6\sigma$). γ is the mean over the thickness series. $L$ is the decay length of an exponential fit to $\Delta P_T$ in the thin-film regime ($\ell_t \lesssim 10$–11σ) and in the thick-film tail. $h^*$ is obtained by interpolation (Sec. II C); its interval comes from Monte Carlo resampling.

| $r_c$ (σ) | $T^*$ | γ ($\varepsilon/\sigma^2$) | $L$ thin (σ) | $L$ thick (σ) | $h^*$ (σ) | 16–84% (σ) | $h^*/r_c$ |
|---|---|---|---|---|---|---|---|
| 3.5 | 0.72 | 0.827 | 1.26 ± 0.10 | 1.3 ± 0.4 | 11.3 | 9.9–11.5 | 3.2 |
| 5.0 | 0.72 | 0.977 | 1.49 ± 0.06 | 1.4 ± 0.7 | 11.7 | 11.1–13.3 | 2.3 |
| 7.5 | 0.72 | 1.051 | 1.75 ± 0.04 | 2.8 ± 0.5 | 14.8 | 14.1–15.7 | 2.0 |
| 7.5 | 0.80 | 0.873 | 1.72 ± 0.14 | 2.4 ± 0.2 | 14.6 | 14.3–14.7 | 1.9 |
| 7.5 | 0.90 | 0.660 | 2.2 ± 0.3 | 2.4 ± 0.7 | 15.0 | 14.1–16.2 | 2.0 |

The surface tensions agree with earlier simulations of the same model at $T^* = 0.72$. At $r_c = 7.5\sigma$, a film of 16.8σ in a box with $b_t = 150.4\sigma$ gave $\gamma = 1.051$ $\varepsilon/\sigma^2$ [6], identical to the present mean. For the STS model, Trokhymchuk and Alejandre report $\gamma = 0.560 \pm 0.002$ (Monte Carlo) and $0.544 \pm 0.018$ (MD) at $r_c = 2.5\sigma$, and list 0.91 at 4.4σ from the MD simulations of Holcomb et al. [4,7]; together with the present 0.827, 0.977 and 1.051 at 3.5, 5.0 and 7.5σ, these values form a single sequence that increases with the cutoff. When the impulsive term is included (their ST model), the tension is larger at the same cutoff: $0.748 \pm 0.012$ at 2.5σ, 0.992 at 4.4σ and 1.042 at 5.5σ (MD) [7]. The comparison confirms that the cutoff dependence of γ in Table II is that of the model, not an artifact of the film thickness.

### E. Effect of temperature

At $r_c = 7.5\sigma$, raising the temperature from $T^* = 0.72$ to 0.90 lowers the surface tension from 1.051 to 0.660 $\varepsilon/\sigma^2$ and lengthens the thin-film decay length from 1.75 to 2.2σ, in line with the broadening of the interface. At a fixed thickness of 11σ the central anisotropy increases with temperature, from 0.0107 at $T^* = 0.72$ to ≈0.014 at 0.80 and ≈0.016 at 0.90 (the last two interpolated with the local decay lengths). The threshold, however, changes little: $h^* = 14.8$, 14.6 and 15.0σ at $T^* = 0.72$, 0.80 and 0.90, values that agree within their uncertainties [Fig. 5(b), Table II]. The larger anisotropy of hot films is compensated by the fact that it is measured against a softer liquid. At $T^* = 0.80$ a film of 14.5σ (three replicas) gives $\Delta P_T = 0.0032 \pm 0.0005$ $\varepsilon/\sigma^3$, at the tolerance itself, which fixes $h^*$ within 14.3–14.7σ.

### F. Effect of the lateral box size

The lateral series tests whether the anisotropy at the film center depends on the box (Fig. 6). For films of 7.75σ at $r_c = 7.5\sigma$, $\Delta P_T$ is $0.0721 \pm 0.0040$, $0.0722 \pm 0.0034$, $0.0677 \pm 0.0008$ and $0.0698 \pm 0.0011$ $\varepsilon/\sigma^3$ at $b_t = 20$, 30, 49.6 and 70σ; a constant value of $0.0686 \pm 0.0006$ fits these with $\chi^2 = 4.3$ for 3 degrees of freedom. For films of ≈14σ, close to the mechanical threshold, the values at $b_t = 20$, 30, 49.6 and 100σ are compatible with a constant of $0.0034 \pm 0.0005$ ($\chi^2 = 2.9$); a term in $1/b_t$ is not significant (1.7 standard errors). At $r_c = 3.5\sigma$ a film of 8.07σ gives $0.0260 \pm 0.0026$ at $b_t = 30\sigma$ (three replicas) and $0.0279 \pm 0.0016$ at 49.6σ. The only difference larger than two standard errors is found for 7σ films at $b_t = 20\sigma$ ($0.120 \pm 0.006$ against $0.1037 \pm 0.0005$), and it is not repeated at 7.75σ; we attribute it to the large scatter between replicas in the smallest box. The surface tension is also independent of $b_t$ within error. Within the precision of the present data the mechanical threshold does not depend on the lateral size, although the value of $\Delta P_T$ at 14σ, which straddles the tolerance, implies that the box could shift $h^*$ by a few tenths of σ.

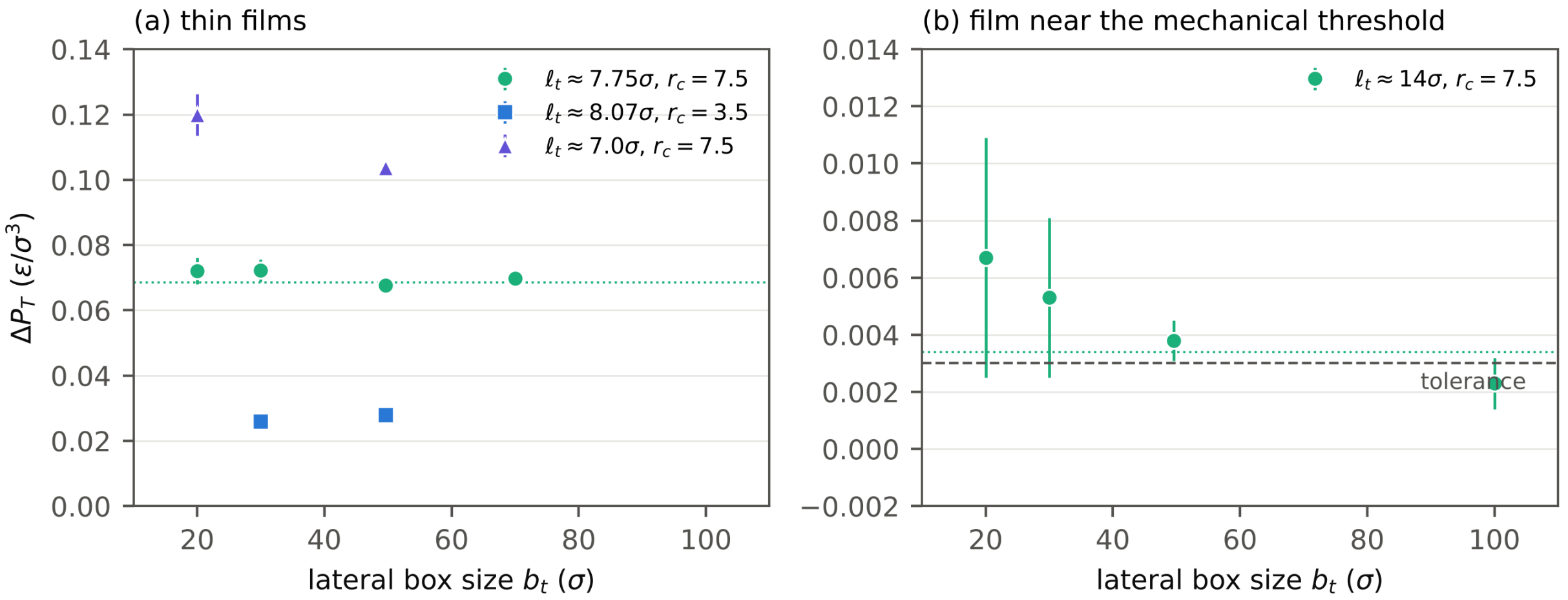

FIG. 6. Central anisotropy against lateral box size for films of constant number of atoms per unit area. (a) Thin films. (b) Films of ≈14σ, close to the mechanical threshold at $r_c = 7.5\sigma$. Dotted lines are weighted means; the dashed line is the tolerance. Films of 7.75σ at $b_t = 100\sigma$ ruptured (Sec. III G) and have no value.

### G. Rupture and the kinetic threshold

Every film that ruptured followed the same sequence: a hole opened, the perforated film turned into a cylinder spanning the cell in one direction and, in most cases, the cylinder broke into a drop (Fig. 2 and Fig. S1). The final state is set by geometry. In the sharp-interface limit, a slab of thickness ℓ in a square cell of side $b_t$ has a larger area than a cylinder of the same volume when $b_t/\ell > \pi$ and than a sphere when $b_t/\ell \gtrsim 3.8$. At $b_t = 20\sigma$ and $\ell = 6\sigma$ the slab and the cylinder have almost the same area (800 against 791 $\sigma^2$) and the sphere is worse (≈890 $\sigma^2$); the three replicas perforated and ended as cylinders, with an energy change of only 0.03 ε per atom that only the morphology could detect. At $b_t = 49.6\sigma$ the cylinder is stable against the Rayleigh–Plateau instability [42] (its circumference, ≈62σ, exceeds the cell length) and only ≈3% larger in area than the drop, and one of the three replicas remained a cylinder for $2 \times 10^7$ steps. The energy released also follows the area: the drop in potential energy for 49,802 atoms at $b_t = 100\sigma$ (0.735 ε per atom) is 1.61 times that for 12,252 atoms at 49.6σ (0.456), against 1.57 predicted from the change in surface area.

The kinetic threshold is summarized in Fig. 7 and Table III. At $b_t = 49.6\sigma$, films of 6σ ruptured and films of 7σ survived $10^5$ τ in all replicas, for each of the three cutoffs; at $r_c = 7.5\sigma$ films of 6.5σ also ruptured. The threshold is therefore the same, between 6 and 7σ, for $r_c = 3.5$, 5.0 and 7.5σ, although the central anisotropy of a 7σ film differs by 60% between the shortest and longest cutoff. The lateral size, which has no measurable effect on $\Delta P_T$, moves the kinetic threshold strongly. At $r_c = 7.5\sigma$ it lies between 6 and 7σ at $b_t = 20\sigma$, between 6.5 and 7σ at 49.6σ, between 7 and 7.75σ at 70σ, and above 7.75σ at 100σ, where a 7.75σ film ruptured after $1.31 \times 10^4$ τ. The same holds for the shortest cutoff: at $r_c = 3.5\sigma$ the 7σ film, which survives $10^5$ τ at $b_t = 49.6\sigma$, ruptured at $b_t = 100\sigma$ in all three replicas, after 120, 120 and 625 τ. At $b_t = 20\sigma$ the 7σ film is the minimum-area shape (800 $\sigma^2$, against 854 for the cylinder and 985 for the drop), so it has no lower-area state to transform into.

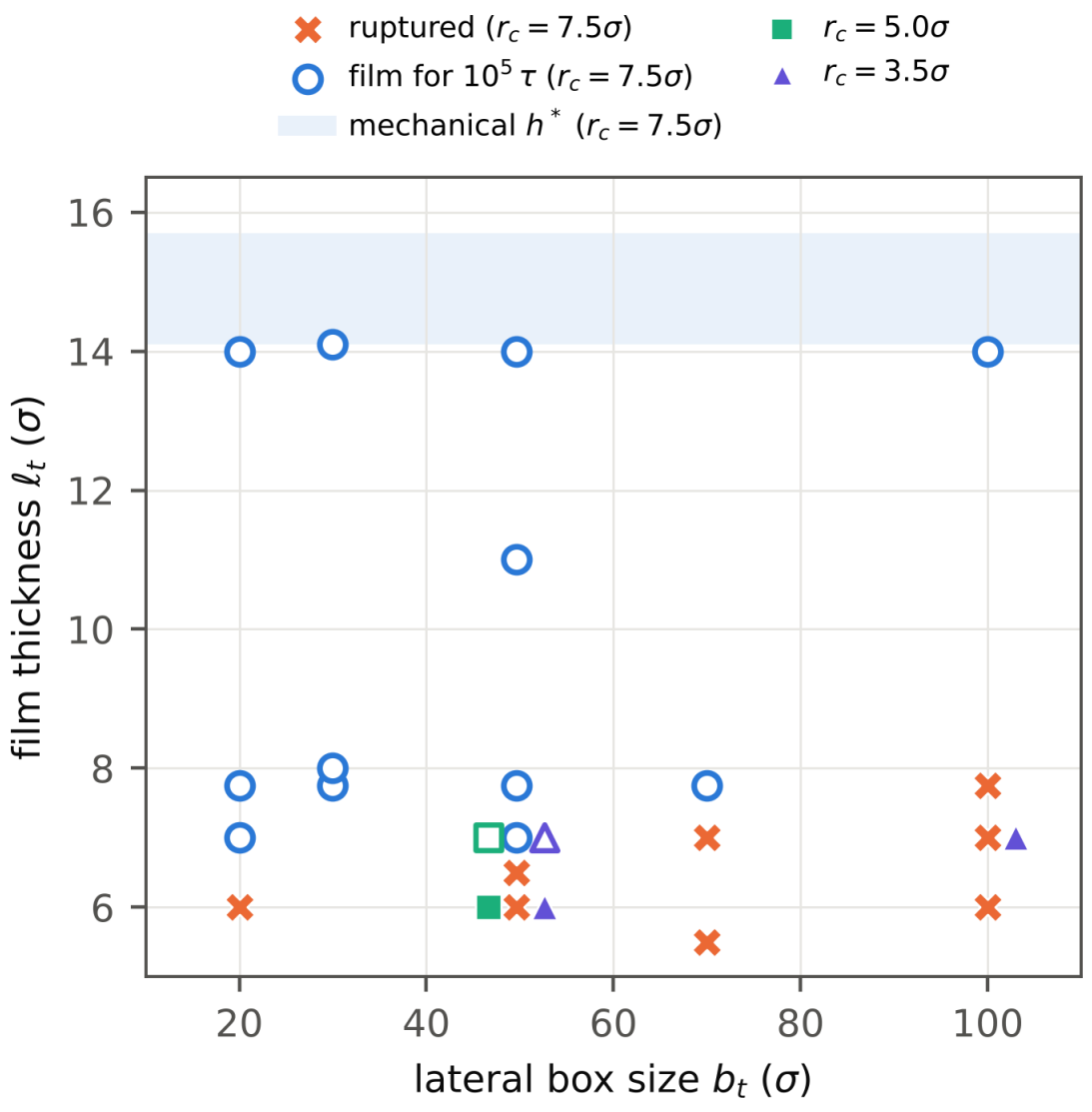

FIG. 7. Kinetic stability of LJ films against lateral box size and thickness. Crosses: rupture in all replicas; open symbols: film for the full $10^5$ τ. Circles and crosses are $r_c = 7.5\sigma$; squares ($r_c = 5.0\sigma$) and triangles ($r_c = 3.5\sigma$, at $b_t = 49.6$ and $100\sigma$) are displaced

horizontally for clarity; filled symbols ruptured, open symbols survived. The shaded band is the mechanical threshold $h^*$ at $r_c = 7.5\sigma$.

TABLE III. Rupture runs at $T^* = 0.72$. Onset times are from the potential energy where a film baseline exists and from the morphology otherwise (ranges). "Film" means that the film survived the $2 \times 10^7$-step equilibration ($10^5$ τ) and the production run. Final states refer to the end of each run; runs were stopped after the final state was reached.

| $r_c$ (σ) | $b_t$ (σ) | $\ell_t$ (σ) | Replicas | Outcome | Onset of rupture (τ) | Final state |
|---|---|---|---|---|---|---|
| 7.5 | 20 | 6.0 | 3 | rupture | < 50 (all three) | cylinder |
| 7.5 | 20 | 7.0 | 3 | film | — | film |
| 7.5 | 49.6 | 6.0 | 3 | rupture | 235, 295, 345 | drop, cylinder, drop |
| 7.5 | 49.6 | 6.5 | 3 | rupture | 8,100, 8,300, 18,750 | drop, drop, cylinder |
| 7.5 | 49.6 | 7.0 | 3 | film | — | film |
| 7.5 | 70 | 5.5 | 1 | rupture | 50–100 | drop |
| 7.5 | 70 | 7.0 | 3 | rupture | 165, 195, 1,890 | drop |
| 7.5 | 70 | 7.75 | 1 | film | — | film |
| 7.5 | 100 | 6.0 | 3 | rupture | 125, 140, 50–100 | drop |
| 7.5 | 100 | 7.0 | 3 | rupture | 90, 90, 95 | drop |
| 7.5 | 100 | 7.75 | 1 | rupture | 13,100 | drop |
| 5.0 | 49.6 | 5.0 | 3 | rupture | < 50 (all three) | drop, cylinder, cylinder |
| 5.0 | 49.6 | 6.0 | 3 | rupture | 405, 1,445, 3,895 | drop |
| 5.0 | 49.6 | 7.0 | 3 | film | — | film |
| 3.5 | 49.6 | 6.0 | 3 | rupture | 660, 1,655, 2,065 | drop, drop, cylinder |
| 3.5 | 49.6 | 7.0 | 3 | film | — | film |
| 3.5 | 100 | 7.0 | 3 | rupture | 120, 120, 625 | drop |

Close to the threshold, rupture is an activated event. The onset times of the three replicas spread over an order of magnitude (165 to 1,890 τ for 7σ at $b_t = 70\sigma$; 405 to 3,895 τ for 6σ at $r_c = 5.0\sigma$), as expected for waiting times of a rare event and unlike the narrow distributions of the thinnest films, which perforated within the first few hundred τ (Fig. 8). At $b_t = 49.6\sigma$ and $r_c = 7.5\sigma$ the mean onset time grows from ≈290 τ at 6σ to ≈11,700 τ at 6.5σ and exceeds the $10^5$ τ window at 7σ; a 0.5σ change in thickness changes the lifetime by a factor of about 40. The cutoff modulates the lifetime without moving the threshold: at 6σ the three onset times at $r_c = 7.5\sigma$ are shorter than all six at $r_c = 5.0$ and 3.5σ, an ordering with probability 1/84 ≈ 0.012 if the nine times came from a single distribution. The same ordering appears for 7σ films at $b_t = 100\sigma$, which ruptured after 90–95 τ at $r_c = 7.5\sigma$ and after 120–625 τ at 3.5σ.

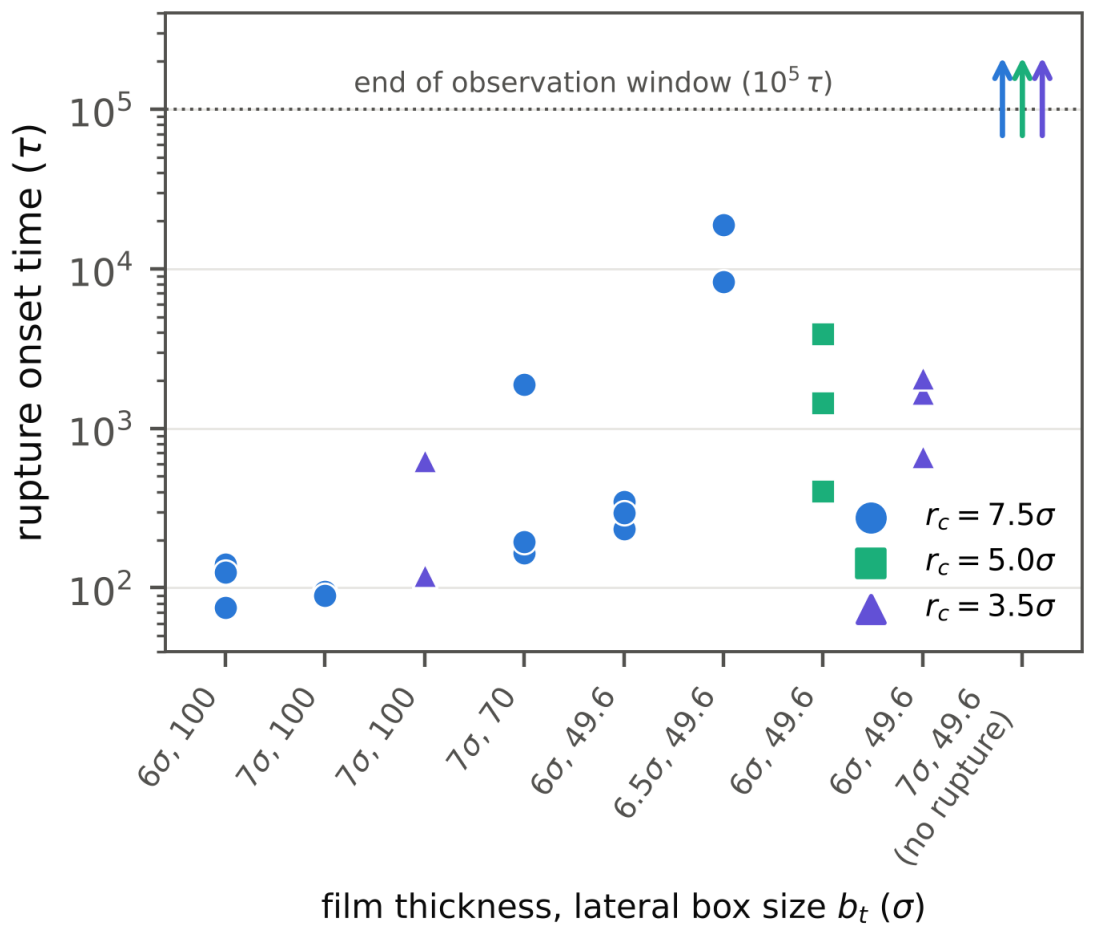

FIG. 8. Onset times of rupture for the films of Table III that ruptured, at $T^* = 0.72$. Labels give thickness and $b_t$. Where only the morphology brackets the onset, the midpoint of the bracket is shown. Arrows: 7σ films at $b_t = 49.6\sigma$, which did not rupture within the observation window for any cutoff.

## IV. DISCUSSION

The two thresholds measure different things, and their opposite responses to $r_c$ and $b_t$ make the difference explicit. The mechanical threshold is a static, local property: it asks whether the liquid at the center feels the presence of the surfaces. It depends on the range of the interactions and on the width of the interfacial region, both of which set how far the stress perturbation of each interface extends, and it does not depend on the box because the mean profile of a stable film is insensitive to its lateral size. The kinetic threshold is a collective property: it asks whether a fluctuation of the film thickness can grow until the film perforates. Such fluctuations are capillary modes whose longest wavelength is set by the box [24,43], which explains why enlarging the box from 49.6 to 100σ shortens the lifetime of a 7σ film from more than $10^5$ τ to about 90 τ. If rupture were a local event with a rate proportional to the film area, the same change would shorten the lifetime only fourfold. The box therefore changes the barrier, not just the number of places where rupture can start. In the classical picture this role is played by a critical wavelength, set by the disjoining pressure, above which thickness modes grow spontaneously [14]; as shown next, that picture does not account for the rupture of these films.

*Relation to the disjoining pressure.* In colloid science, film stability is discussed in terms of the disjoining pressure $\Pi(h)$, the excess of the normal pressure in a film of thickness $h$ over the pressure of the bulk liquid at the same chemical potential [14,15,44,45]. In this section $h$ is the film thickness $\ell_t$, the same quantity that defines the mechanical threshold $h^*$. The central anisotropy $\Delta P_T$ is not Π: it is a tangential and local quantity, whereas Π is normal and is a property of the film as a whole, related to the film tension $\gamma_f = \int(P_N - P_T)dz$, integrated across the whole film, by the Frumkin–Derjaguin relation $d\gamma_f/dh = h\, d\Pi/dh$ [44]. The surface tension reported in Table II is $\gamma = \gamma_f/2$; for a thick film $\gamma_f = 2\gamma_\infty$, where $\gamma_\infty$ is the tension of a single, isolated interface. The two are nevertheless connected, and the connection explains why the surface tension does not see the perturbation. If each surface carried the tangential stress profile of an isolated interface, the anisotropy at the center of a film of thickness $h$ would be the sum of two tails evaluated at $h/2$, $\Delta P_T(h) \approx 2\delta P_T(h/2)$, where $\delta P_T(x)$ is the stress at a distance $x$ inside the liquid of a thick film (Fig. S2). At $r_c = 7.5\sigma$ the tail is 0.044 $\varepsilon/\sigma^3$ at $x = 3.5\sigma$ and 0.029 at 3.9σ, which gives 0.087 for the 7σ film and 0.058 for the 7.75σ film, against 0.104 and 0.068 measured; at $r_c = 5.0$ and 3.5σ the sum gives 67–80% of the measured anisotropy of the films up to 8σ. The anisotropy of thin films is therefore mostly the overlap of two single-interface tails, with a non-additive part of 15–35%. Because tails that nearly add leave the integral of $P_N - P_T$ unchanged, the film tension stays at twice the single-interface value even when the center is strongly perturbed; and because the tails reach further into the liquid for longer cutoffs, $h^*$ grows with $r_c$.

The dispersion part of Π can be bounded independently. For an untruncated $r^{-6}$ attraction, $\Pi = -A/(6\pi h^3)$ and $\gamma_f - 2\gamma_\infty = -A/(12\pi h^2)$, where $A$ is the Hamaker constant, which measures the strength of the dispersion attraction between the two surfaces across the film. For the attractive term $-C/r^6$ of the LJ potential, $C = 4\varepsilon\sigma^6$, and $A = \pi^2 C\rho^2 \approx 27\varepsilon$ at the liquid density $\rho = 0.83\sigma^{-3}$ [46]. For a 7σ film this gives $\Pi \approx -0.004\ \varepsilon/\sigma^3$, 25 times smaller than $\Delta P_T$, and a reduction of the film tension by 0.7%. With the truncated potentials used here the attraction between the two surfaces is weaker still: in the sharp-interface limit it vanishes for $h > r_c$, and at $r_c = 7.5\sigma$ and $h = 7\sigma$ it is 1/40 of the untruncated value. The surface tension of the 7σ film, 0.5% below the mean of the thicker films with a scatter of ±0.6%, is consistent with either estimate. In the linear theory of Vrij [14] a film is unstable to thickness modes longer than $\lambda_c = 2\pi h^2(\pi\gamma/A)^{1/2}$. Even with the untruncated $A$, $\lambda_c = 79\sigma$ for 6σ and 107σ for 7σ, longer than the boxes in which most of these films ruptured; a 6σ film ruptures within ≈300 τ at $b_t = 49.6\sigma$ although no mode longer than $\lambda_c$ fits in the box. Placing $\lambda_c$ at the observed kinetic thresholds would require $A \approx 50$–$130\varepsilon$ and would lower the tension of the 7σ film by 1.2–3.3%, which the measured surface tensions exclude. Moreover, the kinetic threshold is the same at $r_c = 3.5\sigma$, where no dispersion attraction acts across a 6–7σ film. Rupture of these films is therefore not the linear instability of the Vrij picture but an activated event driven by thermal thickness fluctuations, consistent with the broad distribution of onset times; the lateral size presumably enters through the number and the amplitude of these fluctuations rather than through a critical wavelength. The 7σ films at $r_c = 3.5\sigma$ confirm this directly: they survive $10^5$ τ at $b_t = 49.6\sigma$ but ruptured in all three replicas within 625 τ at $b_t = 100\sigma$ (58,000 atoms), as they do at $r_c = 7.5\sigma$. The box dependence of the kinetic threshold therefore does not require any long-range attraction between the surfaces.

These results also explain why, for molten polyethylene, the mechanical criterion gave critical thicknesses two to three times larger than those obtained from visual observation of rupture [13]: the two criteria measure different limits, and only the mechanical one guarantees bulk liquid at the center. Practical consequences for simulations that use free films are listed in the next section.

The study has limitations. The kinetic threshold is defined by a $10^5$ τ observation window and three replicas; a longer window would place it at somewhat larger thicknesses, given the activated character of rupture. The thinnest films perforated during or shortly after the melting of the initial lattice, so their onset times reflect the preparation as well as the film. The production runs of $10^4$ τ may not average slow thickness modes in films close to the kinetic threshold, which is the likely reason why replicas of the 7σ film at $r_c = 3.5\sigma$ scatter by 4.5 times the block error. Finally, only one temperature was explored for rupture.

## V. CONCLUSIONS AND PRACTICAL GUIDELINES

A liquid film can survive without being bulk-like. For LJ films the kinetic limit, below which films rupture within $10^5$ τ, lies between 6 and 7σ at $b_t = 49.6\sigma$ for all three cutoffs and rises with the lateral size of the system; close to it rupture is an activated process whose lifetime changes by more than an order of magnitude for a 0.5σ change in thickness. The mechanical threshold $h^*$, at which the tangential pressure at the center becomes isotropic within 0.003 $\varepsilon/\sigma^3$, is reached only at 11.3σ ($r_c = 3.5\sigma$) to 14.8σ ($r_c = 7.5\sigma$), nearly independently of temperature and lateral size. The central anisotropy is not a disjoining pressure: it is mostly the overlap of the stress tails of the two surfaces, which nearly add and therefore leave the surface tension unchanged, while the dispersion disjoining pressure of these films is too weak to explain their rupture. For simulations that use free films, this suggests five rules:

(1) A film that survives is not necessarily bulk-like at its center. Films between the kinetic limit and $h^*$ (about 2.4–5 nm for argon at $r_c = 7.5\sigma$) should not be used to obtain bulk or single-interface properties without checking against thicker films.

(2) Check the central tangential anisotropy, not the surface tension or the central density: the surface tension shows no thickness dependence, and the relative density deficit is about 16 times weaker than the relative anisotropy and falls below the noise 2–4σ earlier.

(3) Scale the safe thickness with the cutoff: $h^* \approx 11$, 12 and 15σ for $r_c = 3.5$, 5.0 and 7.5σ (tolerance 0.003 $\varepsilon/\sigma^3$), and report it together with $r_c$.

(4) Report the survival of thin films as a lifetime for a given lateral size and run length, not as a critical thickness: enlarging the box from 49.6 to 100σ shortens the lifetime of a 7σ film by more than three orders of magnitude.

(5) Use the Irving–Kirkwood definition of the local pressure, or an estimator of it [31], for thin films: the per-atom-virial profile underestimates the central anisotropy by about a quarter.

The numerical values in these rules are specific to the Lennard-Jones fluid under the conditions studied. The rules themselves rest on mechanisms that are not: the stiffness of a liquid far from its critical point, the near additivity of the stress tails of the two interfaces, and the activated character of rupture. Their quantitative form for molecular liquids, and for interactions of longer range, remains to be established.

## ACKNOWLEDGMENTS

This study received financial support from an internal grant (18379) provided by the Universidad Michoacana de San Nicolás de Hidalgo.

The AI tool mentioned in Sec. II E was also used to draft and edit the text and to check the reference list against publisher records. The author reviewed and verified all the content and takes full responsibility for it.

## DATA AVAILABILITY

The data that support the findings of this article are not publicly available because of the size of the simulation trajectories. The input files, the pressure profiles and the analysis scripts are available from the author upon reasonable request. The numerical results for all the films are tabulated in the Supplemental Material [39].

[1] J. G. Kirkwood and F. P. Buff, The statistical mechanical theory of surface tension, J. Chem. Phys. **17**, 338 (1949).

[2] G. A. Chapela, G. Saville, S. M. Thompson, and J. S. Rowlinson, Computer simulation of a gas–liquid surface. Part 1, J. Chem. Soc., Faraday Trans. 2 **73**, 1133 (1977).

[3] M. J. P. Nijmeijer, A. F. Bakker, C. Bruin, and J. H. Sikkenk, A molecular dynamics simulation of the Lennard-Jones liquid–vapor interface, J. Chem. Phys. **89**, 3789 (1988).

[4] C. D. Holcomb, P. Clancy, and J. A. Zollweg, A critical study of the simulation of the liquid–vapour interface of a Lennard-Jones fluid, Mol. Phys. **78**, 437 (1993).

[5] M. Mecke, J. Winkelmann, and J. Fischer, Molecular dynamics simulation of the liquid–vapor interface: The Lennard-Jones fluid, J. Chem. Phys. **107**, 9264 (1997).

[6] J. L. Rivera and J. F. Douglas, Reducing uncertainty in simulation estimates of the surface tension through a two-scale finite-size

analysis: thicker is better, RSC Adv. **9**, 35803 (2019).

[7] A. Trokhymchuk and J. Alejandre, Computer simulations of liquid/vapor interface in Lennard-Jones fluids: some questions and answers, J. Chem. Phys. **111**, 8510 (1999).

[8] A. Ghoufi, P. Malfreyt, and D. J. Tildesley, Computer modelling of the surface tension of the gas–liquid and liquid–liquid interface, Chem. Soc. Rev. **45**, 1387 (2016).

[9] S. Werth, S. V. Lishchuk, M. Horsch, and H. Hasse, The influence of the liquid slab thickness on the planar vapor–liquid interfacial tension, Physica A **392**, 2359 (2013).

[10] P. Orea, J. López-Lemus, and J. Alejandre, Oscillatory surface tension due to finite-size effects, J. Chem. Phys. **123**, 114702 (2005).

[11] F. Schmitz, P. Virnau, and K. Binder, Determination of the origin and magnitude of logarithmic finite-size effects on interfacial tension: Role of interfacial fluctuations and domain breathing, Phys. Rev. Lett. **112**, 125701 (2014).

[12] J. L. Rivera and J. F. Douglas, Influence of film thickness on the stability of free-standing Lennard-Jones fluid films, J. Chem. Phys. **150**, 144705 (2019).

[13] A. Lemus-Solorio, M. Ramos-Estrada, S. R. Vásquez-García, and J. L. Rivera, Beyond visual observations: establishing the mechanical stability threshold of nanothin polyethylene layers, Membranes **16**, 72 (2026).

[14] A. Vrij, Possible mechanism for the spontaneous rupture of thin, free liquid films, Discuss. Faraday Soc. **42**, 23 (1966).

[15] A. Sheludko, Thin liquid films, Adv. Colloid Interface Sci. **1**, 391 (1967).

[16] E. Ruckenstein and R. K. Jain, Spontaneous rupture of thin liquid films, J. Chem. Soc., Faraday Trans. 2 **70**, 132 (1974).

[17] J. Koplik and J. R. Banavar, Molecular simulations of dewetting, Phys. Rev. Lett. **84**, 4401 (2000).

[18] J.-G. Weng, S. Park, J. R. Lukes, and C.-L. Tien, Molecular dynamics investigation of thickness effect on liquid films, J. Chem. Phys. **113**, 5917 (2000).

[19] J. Becker, G. Grün, R. Seemann, H. Mantz, K. Jacobs, K. R. Mecke, and R. Blossey, Complex dewetting scenarios captured by thin-film models, Nat. Mater. **2**, 59 (2003).

[20] G. Grün, K. Mecke, and M. Rauscher, Thin-film flow influenced by thermal noise, J. Stat. Phys. **122**, 1261 (2006).

[21] R. Fetzer, M. Rauscher, R. Seemann, K. Jacobs, and K. Mecke, Thermal noise influences fluid flow in thin films during spinodal dewetting, Phys. Rev. Lett. **99**, 114503 (2007).

[22] Y. Zhang, J. E. Sprittles, and D. A. Lockerby, Molecular simulation of thin liquid films: Thermal fluctuations and instability, Phys. Rev. E **100**, 023108 (2019).

[23] P. Schofield and J. R. Henderson, Statistical mechanics of inhomogeneous fluids, Proc. R. Soc. London, Ser. A **379**, 231 (1982).

[24] F. P. Buff, R. A. Lovett, and F. H. Stillinger, Interfacial density profile for fluids in the critical region, Phys. Rev. Lett. **15**, 621 (1965).

[25] J. P. R. B. Walton, D. J. Tildesley, J. S. Rowlinson, and J. R. Henderson, The pressure tensor at the planar surface of a liquid, Mol. Phys. **48**, 1357 (1983).

[26] F. Varnik, J. Baschnagel, and K. Binder, Molecular dynamics results on the pressure tensor of polymer films, J. Chem. Phys. **113**, 4444 (2000).

[27] K. Shi, E. R. Smith, E. E. Santiso, and K. E. Gubbins, A perspective on the microscopic pressure (stress) tensor: History, current understanding, and future challenges, J. Chem. Phys. **158**, 040901 (2023).

[28] A. P. Thompson, S. J. Plimpton, and W. Mattson, General formulation of pressure and stress tensor for arbitrary many-body interaction potentials under periodic boundary conditions, J. Chem. Phys. **131**, 154107 (2009).

[29] A. Harasima, Molecular theory of surface tension, Adv. Chem. Phys. **1**, 203 (1957).

[30] J. H. Irving and J. G. Kirkwood, The statistical mechanical theory of transport processes. IV. The equations of hydrodynamics, J. Chem. Phys. **18**, 817 (1950).

[31] J. L. Rivera, Estimating Irving–Kirkwood pressure profiles from per-atom virial and density profiles: a closure for the contour-dependent term in planar liquid films, arXiv:2610.41906.

[32] J. E. Jones, On the determination of molecular fields. II. From the equation of state of a gas, Proc. R. Soc. London, Ser. A **106**, 463 (1924).

[33] B. Smit, Phase diagrams of Lennard-Jones fluids, J. Chem. Phys. **96**, 8639 (1992).

[34] J. Janeček, Long range corrections in inhomogeneous simulations, J. Phys. Chem. B **110**, 6264 (2006).

[35] A. Rahman, Correlations in the motion of atoms in liquid argon, Phys. Rev. **136**, A405 (1964).

[36] A. P. Thompson, H. M. Aktulga, R. Berger, D. S. Bolintineanu, W. M. Brown, P. S. Crozier, P. J. in 't Veld, A. Kohlmeyer, S. G. Moore, T. D. Nguyen, R. Shan, M. J. Stevens, J. Tranchida, C. Trott, and S. J. Plimpton, LAMMPS – a

flexible simulation tool for particle-based materials modeling at the atomic, meso, and continuum scales, Comput. Phys. Commun. **271**, 108171 (2022).
[37] S. Nosé, A molecular dynamics method for simulations in the canonical ensemble, Mol. Phys. **52**, 255 (1984).
[38] W. G. Hoover, Canonical dynamics: equilibrium phase-space distributions, Phys. Rev. A **31**, 1695 (1985).
[39] See Supplemental Material at URL will be inserted by publisher for the complete tables of all simulated films (thickness, density, surface tension, central anisotropy, residuals of the Irving–Kirkwood estimator and partition of the surface tension) and of all rupture runs, the rupture-detection procedure, the validation of the estimator and of the statistical errors, the test of the superposition of single-interface stress tails, the central densities from the full profiles, and the pressures and bulk modulus of the bulk Lennard-Jones liquid.
[40] J. K. Johnson, J. A. Zollweg, and K. E. Gubbins, The Lennard-Jones equation of state revisited, Mol. Phys. **78**, 591 (1993).
[41] M. Thol, G. Rutkai, A. Köster, R. Lustig, R. Span, and J. Vrabec, Equation of state for the Lennard-Jones fluid, J. Phys. Chem. Ref. Data **45**, 023101 (2016).
[42] Lord Rayleigh, On the instability of jets, Proc. London Math. Soc. **10**, 4 (1878).
[43] S. W. Sides, G. S. Grest, and M.-D. Lacasse, Capillary waves at liquid–vapor interfaces: A molecular dynamics simulation, Phys. Rev. E **60**, 6708 (1999).
[44] B. V. Derjaguin, N. V. Churaev, and V. M. Muller, *Surface Forces* (Springer, Boston, 1987).
[45] L. G. MacDowell, J. Benet, N. A. Katcho, and J. M. G. Palanco, Disjoining pressure and the film-height-dependent surface tension of thin liquid films: New insight from capillary wave fluctuations, Adv. Colloid Interface Sci. **206**, 150 (2014).
[46] H. C. Hamaker, The London–van der Waals attraction between spherical particles, Physica **4**, 1058 (1937).